\documentclass[prc,twocolumn,showpacs,preprintnumbers,amsmath,amssymb,nofootinbib]{revtex4-1}

\usepackage{graphicx}
\usepackage{dcolumn}
\usepackage{bm}
\usepackage[mathlines]{lineno}
\usepackage{multirow}
\usepackage{CJK}

\newcommand{\rr} {\boldsymbol{r}}

\begin{document}
\begin{CJK*}{GBK}{song}
\title{ Microscopic description of neutron emission rates in compound nuclei}
\author{Yi Zhu}
\affiliation{State Key Laboratory of Nuclear
Physics and Technology, School of Physics, Peking University,  Beijing 100871, China}
\author{J.C. Pei }
\email{peij@pku.edu.cn}
\affiliation{State Key Laboratory of Nuclear
Physics and Technology, School of Physics, Peking University,  Beijing 100871, China}

\begin{abstract}
\begin{description}
\item[Background]
The neutron emission rates in thermal excited nuclei are conventionally described by  statistical models with a phenomenological level density parameter that
depends on excitation energies, deformations and mass regions.
In the microscopic view of hot nuclei, the neutron emission rates can be determined by the external neutron gas densities without any free parameters.
Therefore the microscopic description of thermal neutron emissions is desirable that can impact several understandings such as survival probabilities of superheavy compound nuclei and neutron emissivity in reactors.

\item[Purpose]
To describe the neutron emission rates in deformed compound nuclei, the external thermal neutron gases are self-consistently obtained based on the Finite-Temperature Hartree-Fock-Bogoliubov (FT-HFB) approach.

\item[Methods]
The Skyrme FT-HFB equation is solved by HFB-AX in deformed coordinate spaces. Based on the FT-HFB approach,  the thermal properties and external neutron gases are properly described with the self-consistent gas substraction procedure.
Then neutron emission rates can be obtained with the densities of external neutron gases. The results
are compared with the statistical model to explore the connections between the FT-HFB approach and the
statistical model.

\item[Results]
The thermal statistical properties of $^{238}$U and $^{258}$U  are studied in detail in terms of excitation energies. The thermal neutron emission rates in $^{238, 258}$U and superheavy compound nuclei $_{112}^{278}$Cn and $_{114}^{292}$Fl are calculated, which agree well with the statistical model by adopting variables from FT-HFB.

\item[Conclusions]
The coordinate-space FT-HFB approach can provide reliable microscopic descriptions of neutron emission rates in hot nuclei,
as well as microscopic constraints on the excitation energy dependence of level density parameters for statistical models.
\end{description}
\end{abstract}

\pacs{24.10.Pa,21.60.Jz,24.60.Dr,21.10.Tg}

\maketitle
\end{CJK*}

\section{Introduction}
The study of highly excited compound nuclei, characteristic of weakening quantum effects such as superfluity and shell effects~\cite{Egido},
is of great interests for heavy ion reactions. In particular, the
survival probabilities of compound superheavy nuclei, as determined by the competition between
thermal neutron emission and fission rates, are related to the choice of cold or hot fusion for the synthesis of new superheavy elements~\cite{superheavy}.
 The description of thermal nuclear
properties are conventionally based on the statistical model~\cite{Weisskopf,Bohr}, which is very successful, however,
invokes a phenomenological level density parameter based on parameterizations of the Fermi gas model.
While the microscopic calculations of level densities are rare~\cite{horoi}.
 The level densities can also be determined by the experimental neutron evaporation spectrum, however, up to very limited excitation energies~\cite{ramirez}.
 Various models of
level densities indeed can cause uncertainties in descriptions of thermal properties and stabilities of hot nuclei.
On the other hand, the compound nuclei can be
described by the Finite-Temperature Hartree-Fock-Bogoliubov (FT-HFB) approach in a microscopic view without any free parameters~\cite{FT-HFB}.
Thus it is very desirable to explore the connections between two pictures: the statistical model and the FT-HFB approach.

In the FT-HFB approach, the compound nuclei can be self-consistently described by quasiparticle excitations due to a finite temperature~\cite{FT-HFB,Egido,khan}.
 In our previous
work, the evolution of fission barriers and neutron gases in terms of excitation energies have been
studied based on the FT-HFB approach~\cite{pei09, pei10},  which can be meaningful for the experimental synthesis of superheavy nuclei~\cite{superheavy}.
 As a further step, in this work we like to study the neutron emission rates in compound nuclei.
In hot nuclei, the thermal equilibrium of neutron evaporation can be obtained by the pressure produced by the external neutron gas (or vapor)~\cite{Kerman}.
Then the neutron emission lifetime is inversely proportional to the density of the neutron gas~\cite{Bonche,Bonche1}.
Presently the FT-HFB approach is adopted to self-consistently take into account the interplay between single-particle motions and pairing in the thermal equilibrium.
For FT-HFB descriptions of hot nuclei, a crucial problem is to treat the continuum contributions to external neutron gases~\cite{Bonche}.
This can be realized by taking the advantages of the coordinate-space HFB approach, by which the nearly uniform neutron gas in large distances can be obtained~\cite{pei09}.
While the conventional HFB approach based on Harmonic Oscillator basis expansion can not treat such kind of surface asymptotics.

In this work, we adopted the HFB-AX solver with finite temperatures to study the properties of hot nuclei in deformed coordinate spaces~\cite{HFB-AX,pei09}. The HFB-AX solver is based on B-spline techniques for axially symmetric deformed nuclei~\cite{teran}.
The FT-HFB equation is solved with a mesh size of 0.6 fm and the order of B-splines is taken as 12.
To study systems that need large coordinate spaces, the hybrid parallel scheme is adopted.
The aim of this paper is to study the neutron emissivity in hot heavy and superheavy nuclei.
In Section \ref{theory},  the relevant FT-HFB formulas and the neutron emission models are given.
In Section \ref{results}, the thermal properties of hot nuclei and calculated neutron emission rates have been discussed.

\section{Theoretical framework}\label{theory}

\subsection{The FT-HFB theory}
The FT-HFB equation in the
coordinate space can be written as~\cite{khan}:
\begin{equation}
  \left[
\begin{array}{cccc}%
 \displaystyle h_{T}(\rr)-\lambda& {\hspace{0.7cm} } \Delta_{T}(\rr) \vspace{2pt}\\
  \displaystyle \Delta_{T}(\rr)& -h_{T}(\rr)+\lambda \\
\end{array}
\right]\left[
\begin{array}{clrr}%
u_i(\rr) \\ \vspace{2pt} v_i(\rr)\\
\end{array}
\right]=E_i\left[
\begin{array}{clrr}%
 u_i(\rr)  \vspace{2pt}\\ v_i(\rr)
\end{array}
\right],\label{HFB}
\end{equation}
where $h_{T}$ and $\Delta_{T}$ are the temperature dependent single-particle hamiltonian and pairing potential, respectively.
For the particle-hole interaction channel, the mostly used Skyrme forces SLy4~\cite{sly4} is adopted.
For the particle-particle channel, the density dependent surface pairing interaction
is used~\cite{mix-pairing}. The pairing strengths are taken as $V_0$=512 MeV fm$^3$  that
can reasonably reproduce the neutron pairing gap of $^{120}$Sn with a pairing cutoff window of 60 MeV~\cite{HFB-AX}.

Compared to the HFB equation at zero temperature, in the FT-HFB equation the density $\rho$ and pairing density $\tilde{\rho}$ have to be modified as~\cite{FT-HFB}
\begin{eqnarray}
\rho(\rr)=\sum_i|u_i(\rr)|^2f_i+|v_i(\rr)|^2(1-f_i),\\
\tilde{\rho}(\rr)=\sum_iv_i(\rr)^{*}(1-2f_i)u_i(\rr),
\label{rho}
\end{eqnarray}
where $f_{i}=\frac{1}{1+e^{E_{i}/kT}}$ denotes the thermal Fermi distribution ($k$T is the temperature). The temperature dependence of other densities
can also be derived straightforwardly.
The entropy $S$ is given by~\cite{FT-HFB}
\begin{equation}
S=-k\sum_{i}[f_i\mathrm{ln}f_i+(1-f_i)\mathrm{ln}(1-f_i)].
\end{equation}
The total free energy $F_{T}$ is given by $F_T$=$E_{T}-ST$, where $E_{T}$ is the intrinsic energy based on
the temperature dependent densities.

\subsection{FT-HFB approach for neutron emission rates}

Based on FT-HFB solutions,  the hot nuclei are in thermal equilibrium and surrounded by external gases.
The external gases, also called vapor, are contributed by unbound continuum states~\cite{Bonche}.
To extract the physical nuclear density distribution, the FT-HFB equation is solved with and without nuclear interactions.
The free neutron gas contribution corresponds to the FT-HFB solutions without nuclear interactions.
This method has been used in calculations of shell corrections by subtracting the unphysical continuum levels based on the Green's function approach~\cite{green}.
In FT-HFB iteration calculations, the total density distributions $\rho_{p,n}(\rr)$ and the gas density distributions $\rho^{gas}_{p,n}(\rr)$ are obtained respectively.
Then the Fermi energies $\lambda_n$ and $\lambda_p$ are self-consistently determined by satisfying the particle number equation,
\begin{equation}\label{subtraction}
\begin{array}{c}
Z =  \displaystyle \int d^3r [\rho_p(\rr)-\rho_p^{gas}(\rr)],\vspace{4pt} \\
N =  \displaystyle \int d^3r [\rho_n(\rr)-\rho_n^{gas}(\rr)].
\end{array}
\end{equation}
In the case of large coordinate spaces and high temperatures, the particle numbers contributed
by the external gases can be nonegligible and the self-consistent subtraction procedure will become important.

Based on the coordinate-space FT-HFB solutions, the uniform neutron gas density distributions can be obtained.
With the uniform neutron gas density $n_{gas}$,  the neutron emission widths $\Gamma_n$ can be given by the nucleosythesis formula~\cite{Bonche}:
\begin{equation}
  \frac{\Gamma_n}{\hbar}=n_{gas}<\sigma v>,
  \label{evap}
\end{equation}
where $\sigma$ is the neutron capture cross section; $<v>$ is the average velocity of particles in the external gas. For simplicity, the neutron cross section $\sigma$ is taken as $\pi R^2$, i.e., the geometrical cross section, where $R$ is the nuclear radius that can be obtained by FT-HFB calculations.
The neutron emission lifetime $\tau$ is related to the width by $\tau={\hbar}/{\Gamma_n}$.
To calculate the statistical average velocity $<v>$, the Fermi occupation number $f(\varepsilon_n)$ in terms of neutron energies $\varepsilon_n$ (i.e., kinetic energies) in the gas is assumed,
\begin{eqnarray}
<v>&=&\frac{\int_{0}^{\infty}f(\varepsilon_n)v(\varepsilon_n)\sqrt{\varepsilon_n}d\varepsilon_n}{\int_{0}^{\infty}f(\varepsilon_n)\sqrt{\varepsilon_n}d\varepsilon_n},\\
f(\varepsilon_n)&=&\frac{1}{1+\text{exp}(\frac{\varepsilon_n-\lambda_n}{kT})},\\
v(\varepsilon_n)&=&\sqrt{\frac{2\varepsilon_n}{m}}.
\end{eqnarray}
Differ from the procedure in Ref. \cite{Bonche}, we explicitly considered the density of the neutron gas, which is actually equivalent to the Bonche's formula~\cite{Bonche}.
Although the FT-HFB approach has been extensively applied to hot nuclei, neutron star crusts and atomic condensates~\cite{Egido,Martin,khan,monrozeau,pei09,pei10a}, practical calculations of neutron emission rates have been rarely undertaken so far. Compared to calculations in Ref. \cite{Bonche} that was based on spherical FT-Hartree-Fock, we extend the applicability of FT-HFB for neutron emission rates in deformed cases as well as superheavy compound nuclei.

\subsection{Statistical model for neutron emission rates}

Statistical models have been widely used to study the thermal neutron emission rates (or widths).
The neutron evaporation width $\Gamma_n$ with an excited energy of $E^{*}$ is given by~\cite{Bohr,wangnan}:
\begin{equation}
\Gamma_n(E^*)=\frac{2mgR^2}{2\pi \hbar^2\rho(E^*)}\int_{0}^{E^{*}-B_n}\varepsilon \rho(E^{*}-B_n-\varepsilon) d\varepsilon
\label{stat}
\end{equation}
where $\rho(E)$ is the level density in terms of excitation energies, and $g=2s+1$ is the spin factor of neutron;
the excitation energy $E^*$ is the difference in the intrinsic energies ($E^*=E_T-E_{T=0}$) from FT-HFB; $B_n$ is the neutron separation energy.
The level density $\rho(E^{*})$ is based on the Fermi gas model, defined as~\cite{egido93}
\begin{equation}
\rho(E^*)= \frac{\sqrt{\pi}\exp(2\sqrt{aE^*})}{12a^{1/4}E^{*5/4}},
\end{equation}
where $a$ is the level density parameter taken as $A/13$ MeV$^{-1}$ as suggested in~\cite{shlomo}. To compare FT-HFB and the statistical model, the variables
$B_n$, $E^{*}$, $R$ in Eq.(\ref{stat}) are taken the same values as in the FT-HFB approach.

\section{Results and Discussions}\label{results}

We have performed calculations for nuclei $^{238}\mathrm{U}$ and $^{258}\mathrm{U}$ at temperatures up to $k$T=2 MeV.
The statistical bulk properties such as, nuclear density distributions, excitation energies, entropy and neutron Fermi energies at different temperatures are discussed in detail.
These statistical properties are directly related to the formula of neutron emission rates.
To see the box size dependence, large coordinate-spaces of 24 fm and 30 fm are adopted.
Then the neutron emission rates are calculated microscopically with comparison with the statistical model.
To study the superheavy compound nuclei, the neutron emission rates in selected $_{112}^{278}$Cn (cold fusion) and $_{114}^{292}$Fl (hot fusion) have also been studied.

 \begin{figure}[t]
  \includegraphics[width=0.50\textwidth]{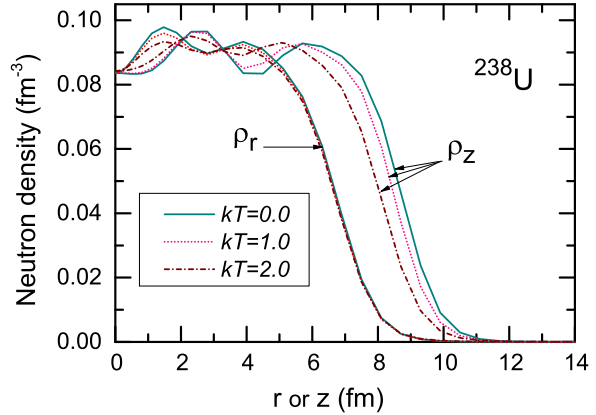}\\
    \caption{(color online) The FT-HFB calculated neutron density distributions of $^{238}\mathrm{U}$  at different temperatures.
    The density profiles $\rho_z$ and $\rho_r$ are shown along the cylindrical $z$-axis and the perpendicular $r$-axis, respectively.    }
     \label{1}
     \end{figure}

  \begin{figure}[t]
  \includegraphics[width=0.9\columnwidth]{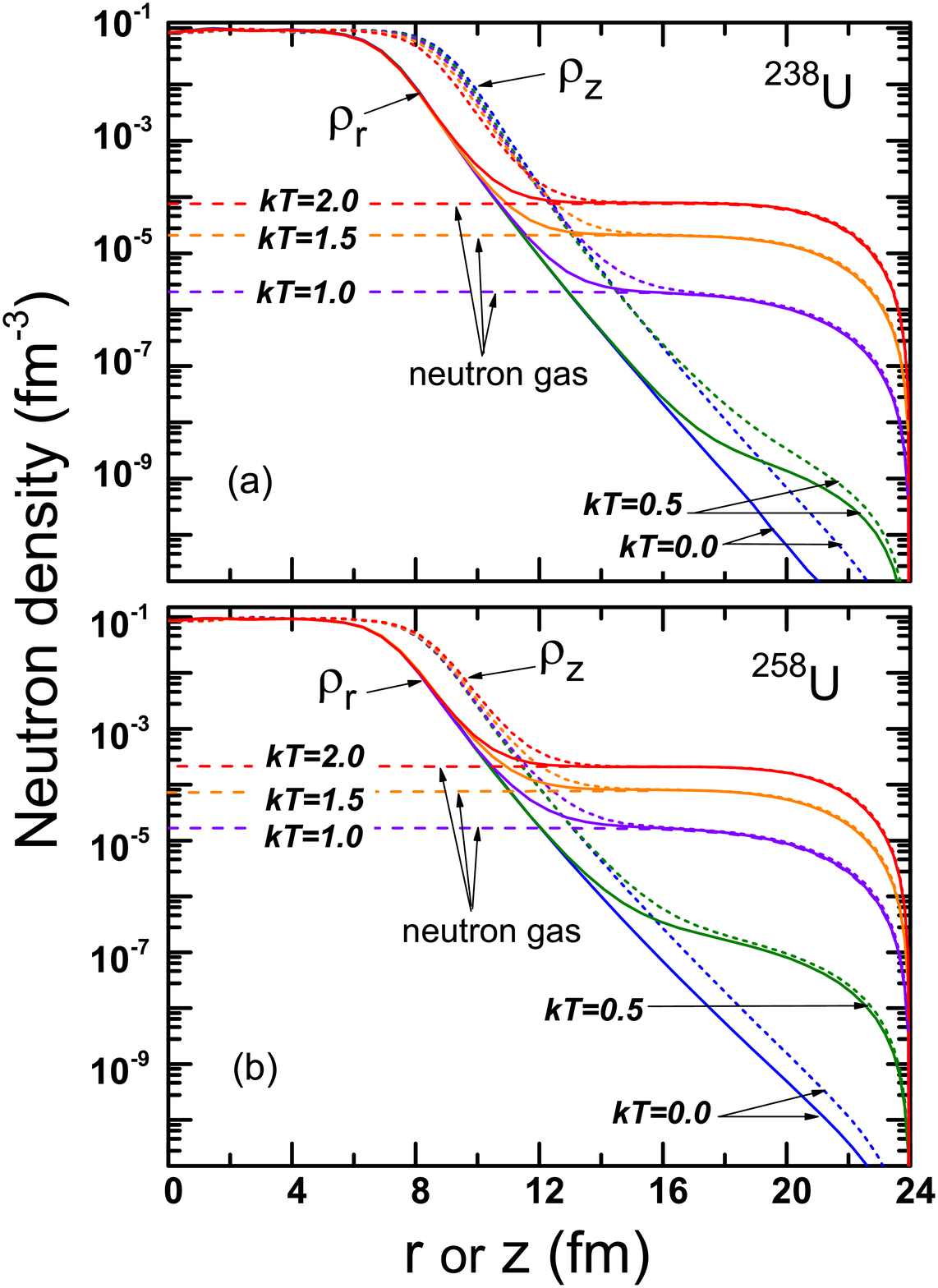}\\
    \caption{(color online) Calculated neutron densities of $^{238}\mathrm{U}$ and $^{258}\mathrm{U}$ at different temperatures. The neutron densities are plotted in logarithmic scale. The external neutron gases are also displayed.  }
     \label{2}
     \end{figure}

\begin{figure}[t]
  \includegraphics[width=0.50\textwidth]{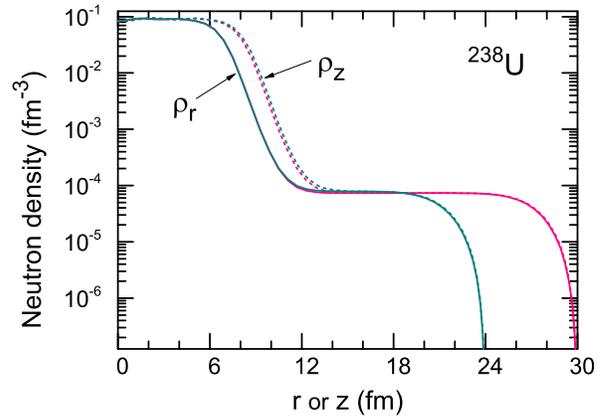}\\
    \caption{(color online) The neutron density distributions (including neutron gas) of $^{238}\mathrm{U}$ at $T$=2.0 MeV,
     obtained by solving FT-HFB equations with box sizes of 24fm and 30fm. }
     \label{boxsize}
\end{figure}

\subsection{Statistical bulk properties of hot nuclei}

Firstly we studied the density distributions of the deformed nucleus $^{238}$U, as displayed in Fig.\ref{1}.
In Fig.\ref{1}, we can see that as temperature increases, (1) the central density fluctuations become weaker,
(2) the density distributions in the $z$-axis direction shrink and the associated shape deformations decrease.
These two observations both agree with the fact that shell effects are diminished as temperatures increase~\cite{Egido,Bonche1}.
Note that the Thomas-Fermi approximation can not take into account such density oscillations~\cite{de96}.
In the case of cylindrical coordinates, the differences between the density profiles $\rho_z(r=0)$ and $\rho_r(z=0)$ actually reflect the surface deformations of axially-symmetric nuclei.
However, the deformation transition to a spherical shape can happen at higher temperatures, depending on different effective nuclear interactions~\cite{Egido,Martin}.

To study the temperature dependent neutron densities including external gases, the neutron density
distributions of $^{238}$U and $^{258}$U are plotted in the logarithmic scale, as shown in Fig.\ref{2}.
As expected, the neutron gas gradually increases with increasing temperatures. At nuclear surfaces,
the neutron gas has a uniform density distribution (values are given in
Table \ref{table1}). Thus the contributions to particle numbers from gases can be easily estimated.
With the same temperature, $^{258}$U always has larger gas densities than $^{238}$U, due to its larger Fermi energies, as shown in Fig.\ref{fig5}.
In addition, in both $^{238}$U and $^{258}$U, one can see that the gas densities increase slower than exponential functions of temperatures.
This behavior has also been shown in compound superheavy nuclei~\cite{pei10}. The proton gases are not shown and their contributions are much smaller due to
the Coulomb potential.

\begin{figure}[htb]
  \includegraphics[width=0.5\textwidth]{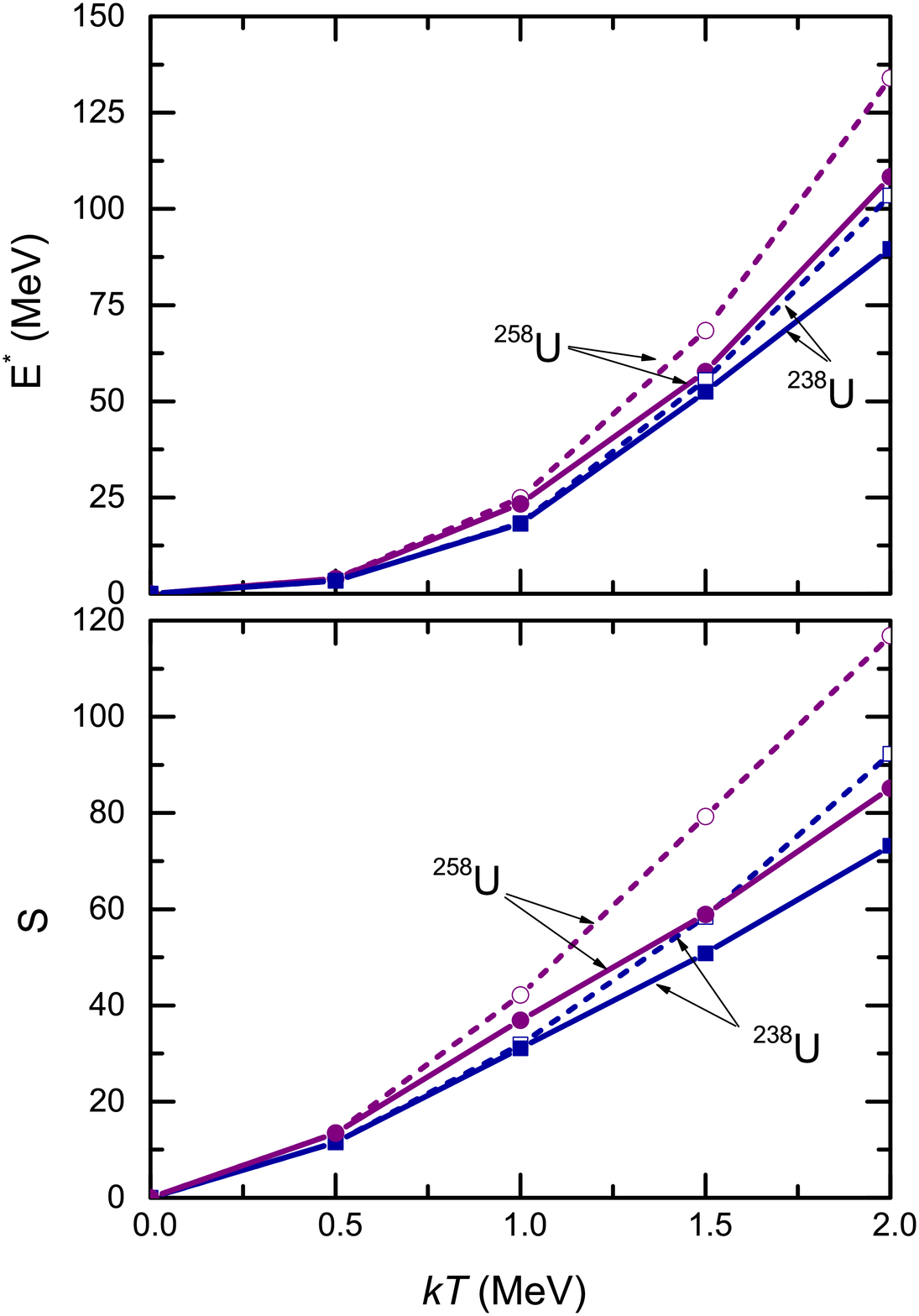}\\
    \caption{(color online) The entropy $S$ and excited energies $E^*$ in $^{238}\mathrm{U}$ and $^{258}\mathrm{U}$ as a function of temperatures. The solid and dashed lines represent results obtained by using and not using the self-consistent gas subtraction procedure, respectively.}
     \label{se}
\end{figure}

\begin{figure}[t]
  \includegraphics[width=0.48\textwidth]{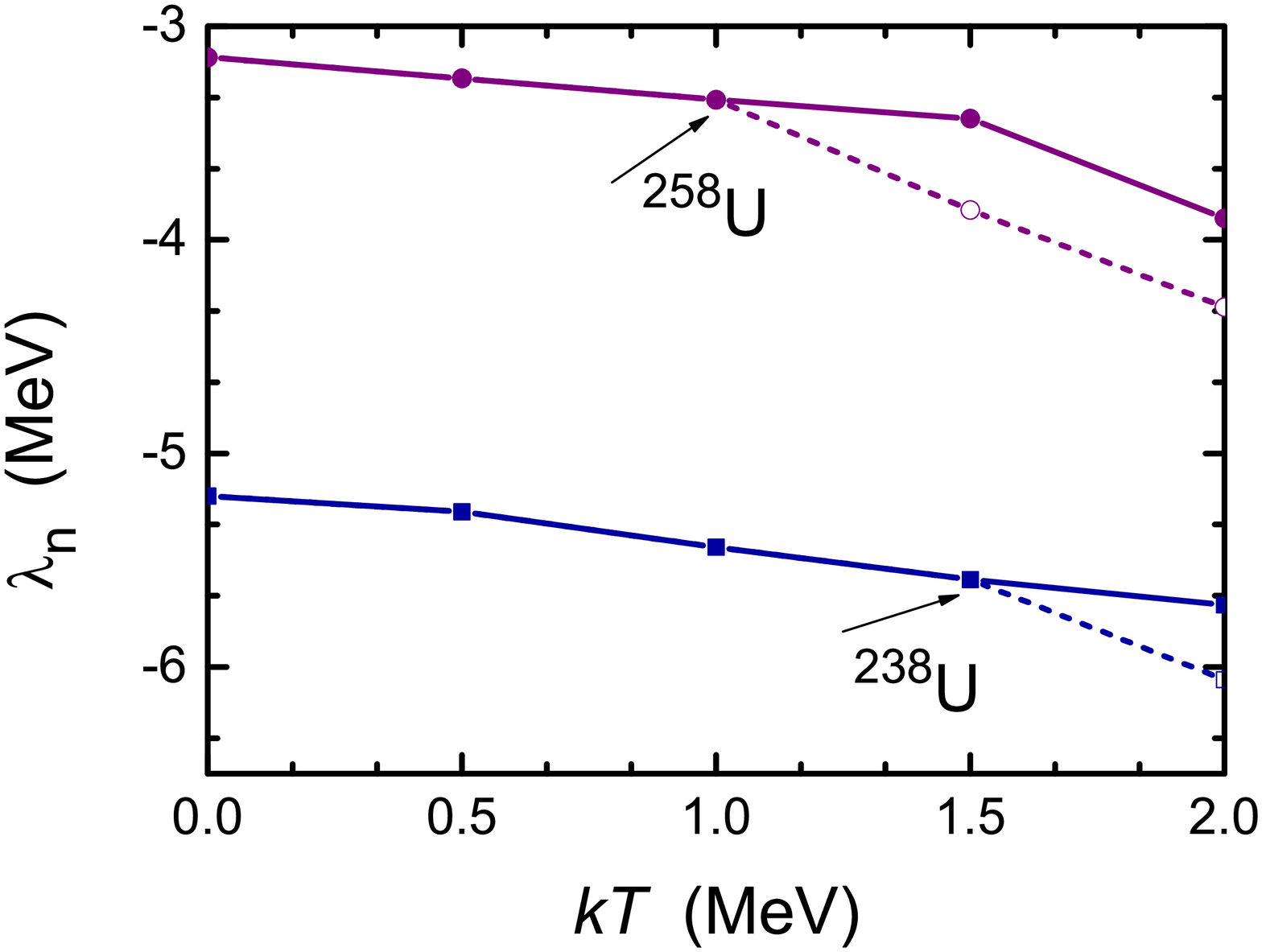}\\
    \caption{(color online) The neutron Fermi energies $\lambda_n$ of $^{238}\mathrm{U}$ and $^{258}\mathrm{U}$ as a function of temperatures. The solid and dashed lines represent results obtained by using and not using the self-consistent gas subtraction procedure, respectively.}
     \label{fig5}
\end{figure}

In Fig.\ref{boxsize}, the neutron density distributions of $^{238}\mathrm{U}$ at $k$T$=2.0\mathrm{MeV}$ are calculated with the box sizes of 24 fm and 30 fm, respectively.
It is important to get convergent neutron gas densities as well as thermal properties by adopting large box sizes~\cite{Bonche1}.
With larger box sizes, the neutron gas contributions become much larger ($\varpropto R_{box}^3$).
 Compared to our earlier work that adopted a box size of 20 fm~\cite{pei09}, the box size adopted in this work is larger and the gas contributions are non-negligible,
particularly in $^{258}$U at $k$T$\geqslant$1.5 MeV.
In our calculations with the self-consistent subtraction procedure (see Eq.\ref{subtraction}), the neutron gas density of 30 fm is slightly less that of 24 fm, within a factor of $5\%$.
This means our calculations are converged and not sensitive to box sizes, in which the subtraction procedure is essential.

To further illustrate the importance of the self-consistent subtraction procedure, the entropy $S$ and excited energies $E^{*}$ as a
function of temperatures are shown in Fig.\ref{se}.
We can see that the gas subtraction in  the neutron-rich $^{258}$U has larger influences than in $^{238}$U.
The subtraction corrections can reduce the entropy and in particular the excitation energies.
Generally, such corrections are significant for temperatures higher than 1.5 MeV.

The Fermi energies (or chemical potentials) of $^{238, 258}$U as a function of temperatures are shown in Fig.\ref{fig5}.
Fig.\ref{fig5} shows that the Fermi energies are weakly dependent on temperatures, and slightly increase within the subtraction approach, implying
increased neutron gas densities. By summarizing Fig.\ref{se} and Fig.\ref{fig5}, we demonstrated that the subtraction procedure is very important,
however, which has not been widely adopted in FT-HFB calculations due to additional computing costs.

\subsection{Neutron emission rates in hot nuclei}

The main aim of this work is to study neutron emission rates based on the neutron gas solutions in the FT-HFB approach.
In this method the neutron emission rates can be calculated microscopically without any free parameters.
The formula have been given in Sec.\ref{theory}. Conventionally the neutron emission rates (or widths) are calculated by
 statistical models that depend on level density parameters.

Table \ref{table1} displays the calculated neutron emission widths as a function of temperatures in $^{238}\mathrm{U}$ and $^{258}\mathrm{U}$, by both the FT-HFB approach
and the statistical model.
In the statistical model, the neutron emission rates are sensitive to the level density parameter $a$.
Firstly we adopted the constant $a=A/13$ that was suggested in ~\cite{shlomo}. We can see that the agreement between the FT-HFB and the statistical model is satisfactory.
For low excitation energies, the statistical model tends to underestimate the emission rates compared to FT-HFB results.
On the contrary, statistical models ($\Gamma(a)$) overestimate the emission rates compared to FT-HFB with high excitation energies.
The discrepancy is due to the utilization of constant level density parameter $a$ that can have temperature dependence to some extent.
In fact, based on the FT-HFB approach, we can extract the level density parameter $a$ by three definitions~\cite{Bonche}: $a=\frac{S}{2T}$, $a=\frac{E^{*}}{T^2}$  and  $a=\frac{S^2}{4E^{*}}$.
Here we adopt $a=\frac{E^{*}}{T^2}$ from FT-HFB solutions to calculate the neutron emission rates by the statistical model ($\Gamma(b)$).
In this case, the agreement between two approaches is also good, but slightly underestimate the widths systematically.
For comparison, $\frac{S^2}{4E^{*}}$ gives the smallest level density parameter and overestimates neutron emission rates.
While $\frac{E^{*}}{T^2}$ gives the largest level density parameter, which is consistent with earlier calculations~\cite{Bonche1}.
This demonstrated the excitation energy dependent level density parameter ($a=\frac{E^{*}}{T^2}$) can be used to connect
the microscopic FT-HFB and the statistical model. Experimental studies of the level density of
$^{238}$U at high excitation energies can provide meaningful examinations, unfortunately, which is known for ${E^{*}}$ only up to 6 MeV~\cite{u238}.

\renewcommand{\arraystretch}{1.3}
\begin{table}[t]
 \caption{\label{table1} The external neutron gas density $n_{gas}$ (fm$^{-3}$)  and the calculated neutron emission widths (in MeV) of $^{238}\mathrm{U}$ and $^{258}\mathrm{U}$.
 The \textrm{Stat-M} widths $\Gamma(a)$ are obtained by the statistical model with the constant level density parameter.   The widths $\Gamma(b)$ are obtained by statistical model with the excitation energy dependent level density parameter. See text for details.
 }

\begin{ruledtabular}
\begin{tabular}{ccccc}
  kT & \multicolumn{2}{c}{FT-HFB}  &  \multicolumn{2}{c}{Stat-M} \\ \cline{2-3}\cline{4-5}
(MeV) & $n_{gas}$ &  $\Gamma$ &  $\Gamma(a)$  &  $\Gamma(b)$  \\  \hline
\multicolumn{5}{c}{$^{238}{\mathrm {U}}$}\\
1.0&$2.07\!\times\! 10^{-6}$ &$3.69\!\times\!10^{-3}$&$1.11\!\times\!10^{-3}$&$1.15\!\times\!10^{-3}$\\
1.5&$2.09\!\times\! 10^{-5}$ &$4.57\!\times\!10^{-2}$&$6.15\!\times\!10^{-2}$&$2.96\!\times\!10^{-2}$\\
2.0&$7.67\!\times\! 10^{-5}$ &$1.94\!\times\!10^{-1}$&$2.56\!\times\!10^{-1}$&$1.55\!\times\!10^{-1}$\\
\multicolumn{5}{c}{$^{258}{\mathrm {U}}$}\\
1.0&$1.67\!\times\! 10^{-5}$ &$3.16\!\times\!10^{-2}$&$1.73\!\times\!10^{-2}$&$1.02\!\times\!10^{-2}$\\
1.5&$7.82\!\times\! 10^{-5}$ &$1.82\!\times\!10^{-1}$&$2.04\!\times\!10^{-1}$&$1.10\!\times\!10^{-1}$\\
2.0&$2.11\!\times\! 10^{-4}$ &$5.70\!\times\!10^{-1}$&$7.71\!\times\!10^{-1}$&$4.10\!\times\!10^{-1}$\\
\end{tabular}
\end{ruledtabular}
\end{table}

\renewcommand{\arraystretch}{1.3}
\begin{table}[htb]
 \caption{\label{table2} Similar to Table \ref{table1}, but for $_{112}^{278}$Cn and $_{114}^{292}$Fl.}

\begin{ruledtabular}
\begin{tabular}{ccccc}
  kT & \multicolumn{2}{c}{FT-HFB}  &  \multicolumn{2}{c}{Stat-M} \\ \cline{2-3}\cline{4-5}
(MeV) & $n_{gas}$ &  $\Gamma$ &  $\Gamma(a)$ &  $\Gamma(b)$  \\  \hline
\multicolumn{5}{c}{$_{112}^{278}$Cn}\\
1.0&$5.53\!\times\! 10^{-7}$ &$1.09\!\times\!10^{-3}$&$1.36\!\times\!10^{-4}$&$5.14\!\times\!10^{-4}$\\
1.5&$8.84\!\times\! 10^{-6}$ &$2.14\!\times\!10^{-2}$&$2.76\!\times\!10^{-2}$&$2.04\!\times\!10^{-2}$\\
2.0&$3.76\!\times\! 10^{-5}$ &$1.06\!\times\!10^{-1}$&$2.21\!\times\!10^{-1}$&$1.21\!\times\!10^{-1}$\\
\multicolumn{5}{c}{$_{114}^{292}$Fl}\\
1.0&$1.15\!\times\! 10^{-6}$ &$3.69\!\times\!10^{-3}$&$5.36\!\times\!10^{-4}$&$9.04\!\times\!10^{-4}$\\
1.5&$1.51\!\times\! 10^{-5}$ &$3.79\!\times\!10^{-2}$&$3.13\!\times\!10^{-2}$&$2.64\!\times\!10^{-2}$\\
2.0&$5.45\!\times\! 10^{-5}$ &$1.58\!\times\!10^{-1}$&$2.51\!\times\!10^{-1}$&$1.48\!\times\!10^{-1}$\\
\end{tabular}
\end{ruledtabular}
\end{table}

Table \ref{table2} displays the calculated neutron emission rates of superheavy compound nuclei $_{112}^{278}$Cn and $_{114}^{292}$Fl, which are
typical cold fusion and hot fusion compound nuclei, respectively.
Again the neutron widths given by FT-HFB and by the statistical model with the excitation energy dependent level density parameter ($a=\frac{E^{*}}{T^2}$) agree very well.
However, the agreement between FT-HFB and the statistical model with the constant level density parameter is less satisfactory, compared to Table \ref{table1}.
The statistical model with the constant $a$ gives very small widths ($\Gamma(a)$) at low temperatures compared to FT-HFB results.
This indicates that the constant level density parameter may not be suitable for the
superheavy mass region.
The neutron emission rates of  $_{112}^{278}$Cn (cold fusion) are smaller than that of $_{114}^{292}$Fl (hot fusion) with the same temperature.
Or say the cooling of $_{114}^{292}$Fl is more favorable by neutron evaporations than $_{112}^{278}$Cn.
This is mainly due to its smaller neutron separation energy.
The microscopic FT-HFB calculations of neutron emission rates can be useful for the synthesis of superheavy nuclei.
The level density connection between FT-HFB and statistical models as pointed out in this work can also be useful for fission studies.

\section{Summary}
In summary, the Finite-Temperature HFB approach has been applied to calculations of neutron emission rates of compound nuclei in the microscopic view.
The FT-HFB equation is solved in the deformed coordinate-space so that the continuum contributions to
external neutron gas can be precisely obtained. We have demonstrated that the self-consistent gas subtraction
is important to properly describe statistical properties of hot nuclei.
The FT-HFB calculated neutron emission rates of compound deformed $^{238, 258}$U and superheavy nuclei agree well with the statistical model.
Furthermore,  we demonstrated that by adopting inputs from FT-HFB (an excitation energy depent level density parameter, excitation energies and neutron separation energies), the statistical model
can be connected to the microscopic FT-HFB approach. Our approach can be useful for the synthesis of superheavy nuclei and
further FT-HFB study of thermal fission rates are in progress.

\section{acknowledgments}
Useful discussions with W. Nazarewicz and F.R. Xu are gratefully
acknowledged.
This work was supported by the Research Fund for
the Doctoral Program of Higher Education of China (Grant
No. 20130001110001), and the National Natural Science Foundation of China under Grants No.11375016, 11235001.
We also acknowledge that computations in this work were performed in the Tianhe-1A supercomputer
located in the Chinese National Supercomputer Center in Tianjin.


\end{document}